\documentclass[namedreferences]{SolarPhysics}
\usepackage[optionalrh]{spr-sola-addons} 
\usepackage{graphicx}        
\usepackage{color}           
\usepackage{url}             

\newcommand{\etal}{{\it et al.}}
\newcommand{\eg}{{\it e.g.}}

\newcommand{\mnras}{  {\it Mon. Not. Roy. Astron. Soc.}}

\newcommand{\pasj}{   {\it Pub. Astron. Soc. Japan}}

\newcommand{\kfnt} [4]{ #1, {\it Kinematika i Fizika Nebes. Tel [Kinematics and Physics of
Celestial Bodies]}, {\bf  #2}, No. #3, #4.}
\newcommand{\arep} [3]{#1, {\it Astron. Rep.}, {\bf  #2}, #3.}
\newcommand{\aeta}[3]{  #1, {\it Astron. Astrophys.}, {\bf  #2}, #3.}
\newcommand{\aetas}[3]{ #1, {\it Astron. Astrophys. Suppl. series}, {\bf  #2}, #3.}
\newcommand{\aspj}[3]{  #1, {\it Astrophys. J.}, {\bf  #2}, #3.}
\newcommand{\sph}[3]{   #1, {\it Solar Phys.}, {\bf  #2}, #3.}

\newcommand{\monras}[3]{#1, {\it Mon. Not. Roy. Astron. Soc.}, {\bf  #2}, #3.}
\begin{document}
\begin{article}

\begin{opening}

\title{Stokes Diagnostis of 2D MHD-simulated Solar
Magnetogranulation}

\author{V.A.~\surname{Sheminova}
       }
\runningauthor{Sheminova} \runningtitle{Stokes Diagnostics of
2D MHD-simulated Solar Magnetogranulation}

   \institute{Main Astronomical Observatory, National Academy of
Sciences of Ukraine, 27 Akademika Zabolotnoho St., Kyiv,
03680~Ukraine
                     email: \url{shem@mao.kiev.ua}
             \\
             }

\begin{abstract}

 We study the properties of  solar magnetic fields on scales
less than the spatial resolution of solar telescopes. A
synthetic infrared spectropolarimetric diagnostics based on a
2D MHD simulation of magnetoconvection is used for this. We
analyze two time sequences of snapshots that  likely
represent two regions of the network fields with their
immediate surrounding on the solar surface with  the unsigned
magnetic flux density of 300~G and  140~G. In the first region
we find from probability density functions of the magnetic
field strength that the most probable field strength at $\log
\tau_5=0$ is equal to 250~G. Weak fields ($B<500$~G) occupy
about 70\%  of  the  surface, while stronger fields
($B>1000$~G) occupy only 9.7\% of the surface. The magnetic
flux is $-28$~G and its imbalance is $-0.04$. In the second
region, these parameters are correspondingly equal to 150~G,
93.3\%, 0.3\%, $-40$~G, and $-0.10$. We estimate the
distribution of line-of-sight velocities on the surface of
$\log\tau_5=-1$. The mean velocity is equal to 0.4~${\rm
km~s^{-1}}$ in the first  simulated region. The averaged
velocity  in the granules is $-1.2$~${\rm km~s^{-1}}$ and  in
the intergranules is $2.5~{\rm km~s^{-1}}$.  In the second
region, the corresponding values of the mean velocities are
equal to 0, $-1.8$, 1.5~${\rm km~s^{-1}}$.  In addition we
analyze the asymmetry of synthetic Stokes-V profiles of the
Fe~I 1564.8~nm line. The mean values of the amplitude and
area asymmetry do not exceed 1\%.  The spatially smoothed
amplitude asymmetry is increased to 10\% while the area
asymmetry is  only slightly varied.

\end{abstract}
\keywords{Magnetic fields, Photosphere; Velocity Fields,
Photosphere; Granulation; Spectral Line, Intensity and
Diagnostics}
\end{opening}

\section{Introduction}
     \label{S-Introduction}

Despite a long history of the solar magnetic field studies
and enormous number of publications,  many problems remain
unsolved. One of these problems concerns solar small-scale
magnetic fields  in the photosphere outside sunspots.  The
determination of the fine structure of magnetic elements is
difficult because of their very weak magnetic flux and
insufficient  resolution. Present-day spectropolarimetric
observations of absorption lines which provide  basic
information about the magnetic field structure have a spatial
resolution  not better than a few tenths of arcsec. To
resolve separate small magnetic features in the photosphere
and to obtain information about their properties and
structure,  we need a resolution better than
$0^{\prime\prime}$.1. It is found  that the small-scale
magnetic elements should be smaller than 50-100~km
\cite{Dominguez03, Khom03}. Furthermore,
\inlinecite{Cameron05} have shown that any magnetic element
should be structured  down to the turbulent magnetic
diffusivity length-scales of 10~km. Such small structures
cannot be resolved with the present-day telescopes. But it is
possible  to improve  our understanding of the nature of
solar small-scale fields with the use of the self-consistent
models of solar photosphere regions. These models are
constructed on the basis  of realistic physics with the use
of the equations of radiation magnetohydrodynamics (MHD).
Such time-dependent MHD models allow us to study in detail
the evolution, dynamics, and structure of magnetic elements.
The spatial resolution of the numerical simulation is  a few
grid points  corresponding to a few tens of kilometers
\cite{Sanchez06}.

A detailed review of realistic MHD simulations of solar
magnetoconvection can be found in \inlinecite{Schussler03}
and  \inlinecite{Steiner07}. Three-dimensional (3D) models of
the solar magnetoconvection were first considered in  the
pioneering work of \inlinecite{Nordlund85}. Later
\inlinecite{Cattaneo99} has studied the local dynamo
mechanism using idealized 3D simulation of thermal
magnetoconvection in the Boussinesq approximation in the
quiet photosphere. Realistic 3D MHD models were further
developed in \citeauthor{Stein00}, (\citeyear{Stein00,
Stein06}); \inlinecite{Carlsson04}; \inlinecite{Vogler05};
\inlinecite{Steiner08}. The 3D MHD simulation are
complemented  by two-dimensional (2D) MHD simulation (\eg,
\opencite{Deinzer84}; \opencite{Brandtg95};
\citeauthor{Atroshchenko96a}, \citeyear{Atroshchenko96a,
Atroshchenko96b}; \opencite{Grossmann-Doerth98};
\opencite{Gadun00}).  The analysis of 2D simulations of
convection \cite{GadunSo99, Ploner99} showed that 2D models
reproduce many features of 3D convection, although  cannot
represent realistic flow patterns \cite{Asplund00}. However,
2D MHD models quite adequately reproduce small-scale
phenomena and are useful in studying the properties of
magnetic elements (\eg, \opencite{Grossmann-Doerth98};
\opencite{Gadun99a}).

The 2D MHD models by \inlinecite{Gadun99a} first presented
the evolution of magnetogranulation of mixed-polarity  solar
regions  over 2 hours of solar time. Based on these models
\inlinecite{Sheminova00} have considered formation and
destruction of fluxtubes and have obtained the observable
signatures of convective collapse in Stokes-V profiles. These
signatures were  confirmed by spectropolarimetric
observations of \inlinecite{Bellot01} and
\inlinecite{Nagata08}.  A clear relation was found between
field strength and magnetic inclination \cite{Gadun01}. This
result is in qualitative agreement with observations
\cite{Lites98}. It was found that the predominantly
horizontal weak fields are located in the photospheric layers
of granules  \cite{Gadun00}.  This fact was confirmed by new
Hinode observations \cite{Lites08}. An example of flux
recycling events in this simulation was discovered by
\inlinecite{Ploner01a}. Later,  3D~MHD simulation  by
\inlinecite{Stein02} confirmed that  a local short-period 
recirculation can arise near the surface under the conditions
of strong stratification and asymmetric convective flows.
This mechanism can provide a high rate of flux emergence
through the solar surface.

The 2D MHD models \cite{ Gadun99a} were  also used in the
Stokes diagnostics of small-scale magnetic fields
(\citeauthor{Shem99}, \citeyear{Shem99, Shem03, Shem04,
Shem05}; \opencite{Sheminova00}; \opencite{Ploner01b}). In
this paper a Stokes diagnostics  of simulated 2D
magnetogranulation with two different  magnetic flux
densities is presented.

As to the terminology, the term "magnetogranulation" means
granulation  in the presence of magnetic field. The compact
magnetic elements  are similar to sheet-like structures
extending along the lanes ("fluxsheets") in the 2D Cartesian
description. We use for them the traditional definition
"fluxtubes".  To avoid confusion in defining the magnetic
field characteristics we  use the following: $B$ is the
magnitude of the magnetic vector, i.e., the  field strength;
$<B>$ is the average field strength over given region; $B_z$
is the vertical component of the magnetic vector; $<|B_z|>$
is the unsigned (absolute) flux density; $<B_z>$  is the flux
density. The magnetic flux $F$ is the area integral over the
field vector when the line of sight is perpendicular to the
area of integration, i.e., it is the longitudinal magnetic
flux.

The paper is arranged as follows: Section 2 shortly describes
the 2D MHD  simulations used in our analysis. The spectral
synthesis of Stokes profiles is considered in Section 3. The
results of Stokes diagnostics are presented in Section 4, and
Section 5 gives our conclusions.

\section{Numerical 2D MHD Models of Solar Magnetogranulation}
\label{2D MHD models}

The complete system of radiative magnetohydrodynamical
equations is used for the 2D MHD simulation of the upper
solar convection zone and the photosphere \cite{Gadun99a}. We
adopt the approximation that the medium is compressible,
partially ionized, stratified by gravity, and coupled with
the radiation.  The simulations start with a 2D
hydrodynamical (HD) model of the solar granulation obtained
earlier by \inlinecite{GadunSo99}. The code for the
integration of the MHD equations represents an extension of
the radiation-hydrodynamic code that is described in detail
by \inlinecite{GadunSo99}. Note that the radiative transfer
is treated in the grey approximation. The magnetic field is
described by the vector potential, so that the divergence of
magnetic field strength is always zero in the simulated
region.

\begin{figure} 
 \centerline{\hspace*{0.015\textwidth}
               \includegraphics[width=0.45\textwidth,clip=]{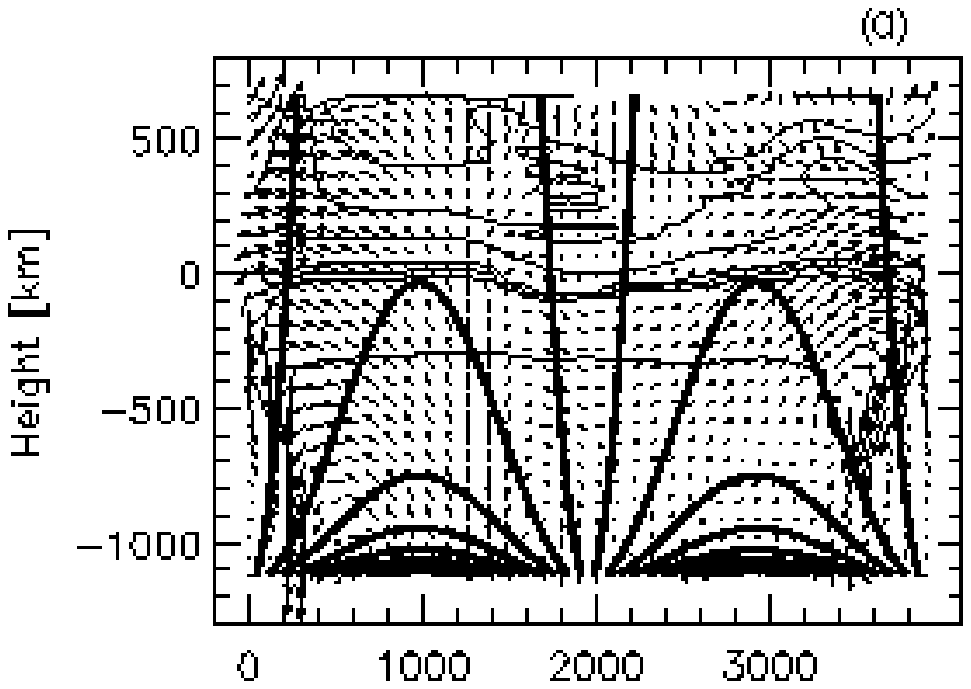}}
 \centerline{\hspace*{0.015\textwidth}\vspace{-0.01\textwidth}
              \includegraphics[width=0.65\textwidth,clip=]{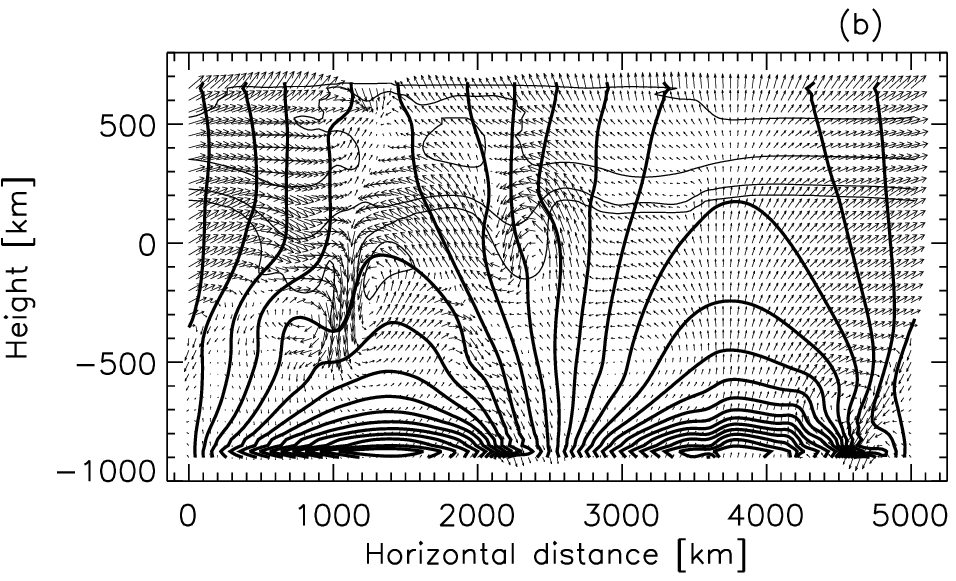}
             }
 \vspace{0.015\textwidth}  

              \caption{(a): snapshot of the
               initial HD model and the initial
configuration of magnetic field in the first MHD simulation
with the spatially averaged magnetic field strength $<$$B$$>$
= 54~G. (b): the same but for $t= 30$~s  in  the second MHD
simulation with $<$$B$$>$ = 1.6~G.  Thick curves: magnetic
field lines. Thin curves: isotherms from top to bottom
indicate the temperature of 4000, 5000, 6000, 7000, 8000,
9000, 10000, and 12000~K. Arrows: velocity field.
             }
\label{init_16}
\end{figure}

In the first simulation  the computational domain spans
3920~km horizontally and 1820~km vertically, with the  grid
step of 35~km corresponding to $112\times$52
grid-points.  The atmosphere is extended from
1135~km below to 685~km above the geometrical height $h=0$ or
$\tau_R=1$, $\tau_R$ being the Rosseland optical depth
(Figure~\ref{init_16}a). This simulation ran for 2 hours of
the solar time.

In the second simulation the grid size of the computational
domain  is $202\times64$ grid-points (or $5050\times
1600$~km, Figure~\ref{init_16}b). The corresponding grid step
is 25~km. The atmosphere covers 900~km below and 700~km above
the level $h=0$. The total simulation time is 4.5~hours.

The side boundaries are periodic while the upper and lower
boundaries are open. We adopt $B_x=0$ and $dB_z /dz =0$ at
the upper and lower boundaries, where  $B_x$ is  the
horizontal component of the magnetic field. The configuration
of the initial magnetic field is bipolar (Figure~\ref{init_16}).
The  total initial field flux over the computational box is zero.
The mean initial magnetic field strength $<B>$ over the
entire computational domain in the first and second simulations is  54
 and 1.6~G. Processing and amplification
of the initial field is resulted in additional magnetic unsigned
flux, which is varied with time. The mechanisms of the magnetic
field  growth were considered by \inlinecite{Gadun00}.

\noindent

\begin{figure}
 \centerline{ \includegraphics[width=0.85\textwidth]{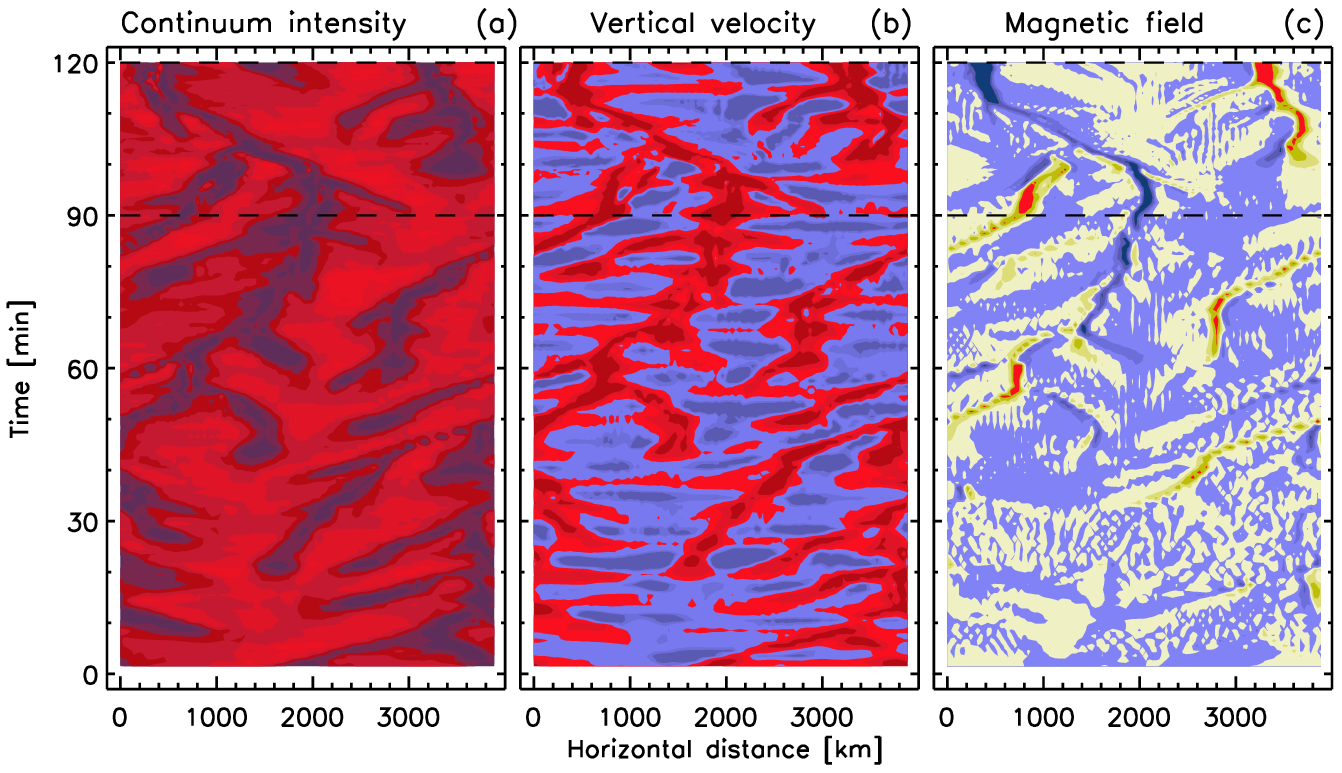}}
 \centerline{\hspace*{0.068\textwidth} \includegraphics[width=0.75
             \textwidth]{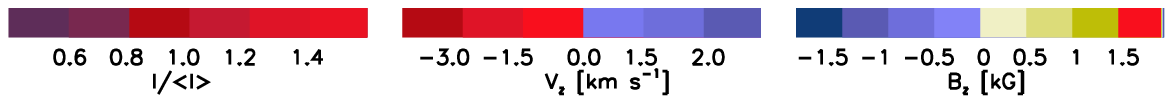}
              }
           \caption {The maps showing the spatial-temporal evolution
of the continuum intensity at $\lambda=500$~nm (a), the
vertical component of velocity field (b), and  magnetic field
(c) at the level $\log \tau_R=0$ in the first MHD simulation.
The horizontal axis indicates position on the Sun, while time
increases  upwards. Positive values of the velocity shown in
blue correspond to upflows  while  downflows are indicated in
red. The different scales  for positive and negative values
of vertical magnetic component $B_z$ indicate negative and
positive polarity  field. Dashed lines mark the selected
region for Stokes diagnostics with absolute flux density of
300~G (300-MHD).
              }
\label{m_54}
\end{figure}

The initial MHD model includes the initial magnetic field
superposed on the  HD model of solar granulation. A transient
of the order of 30 min is needed for the magnetic field to be
consistent with the HD structure. In general, the simulation
may be arbitrarily divided into three stages.
Figure~\ref{m_54} shows the evolution of the brightness, of
the vertical velocity, and of the vertical magnetic field in
the first simulation at $\tau_R=1$.  The initial period
(0--20 min) is determined mainly by the initial conditions:
the field lines are forced out  by convective motions to the
region of downflows. The field lines are partially
reconnected and the local field concentrations are
dissipated. In the period, $t=20$--30~min,  no
clearly-defined features of the magnetic field is present.
This can be seen from the $B_z$-distribution
(Figure~\ref{m_54}c). Therefore the magnetic field has a
chaotic (turbulent) distribution up to $t=30$~min. Later on
the field becomes structured: it is enhanced in intergranular
lanes; new thin flux tubes begin to form at the moment of the
granule fragmentation;  fields of different polarities are
separated; the field in strong flux tubes tends to be
vertical, while it is more horizontal above granules. The
structural changes of the magnetic field, the brightness
field, and the vertical velocity field are similar in
appearance. After $t=50$~min the bright points appear within
the intergranular lanes (Figure~\ref{m_54}a) corresponding to
the regions with compact strong predominantly vertically
orientated magnetic field (Figure~\ref{m_54}c) and strong
downfllows (Figure~\ref{m_54}b).

The spatial-temporal changes of  the magnetogranulation in
the first simulation (Figure~\ref{m_54}) can be considered as
a filtergram, a Dopplergram, and a magnetogram on the solar
surface obtained with a slit which cuts 35~km strips at each
instant of time every 30 or 60~sec.  By analogy with
observations we  can use a sample of snapshots for Stokes
diagnostics. We select the 30~min sample of snapshots in the
$t=90.5$--120~min interval marked by dashed lines in
Figure~\ref{m_54}. It contains 56 snapshots. Four of them are
taken at $t=90.5$--93.5~min every 1~min, whereas the timestep
of the rest is 30 sec.  $B_z$ in this sample varies from
$-2870$ to 2350~G. Several bright points (BPs) are present in
this simulated  region. The mean unsigned field strength
$<B>$ at the optical depth $\log \tau_R=0$ over all these
snapshots is 498~G. The mean unsigned longitudinal field
strength (or absolute flux density) $<|B_z|>$ amounts to
300~G at $\log \tau_R=0$. The large difference between $<B>$
and $<|B_z|>$ can be explained by the presence of the
relatively weak almost horizontally oriented  fields which
fill the  granular surface. The 30 min sample of snapshots
with absolute flux density of 300~G will be called hereafter
300-MHD.
\begin{figure}
 \centerline{\includegraphics[width=0.97\textwidth]{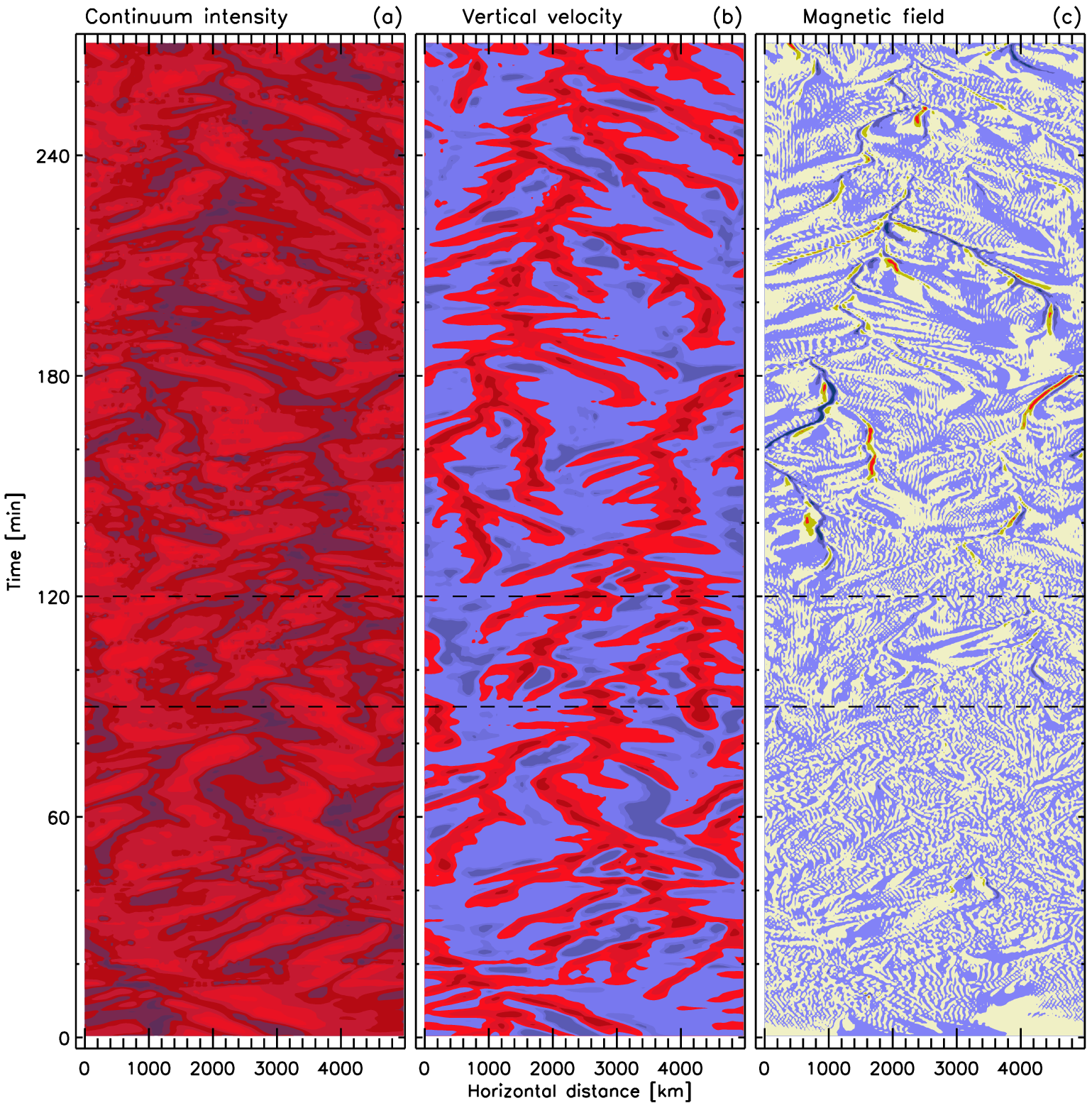}}
 \centerline{\hspace*{0.068\textwidth} \includegraphics[width=0.9
             \textwidth]{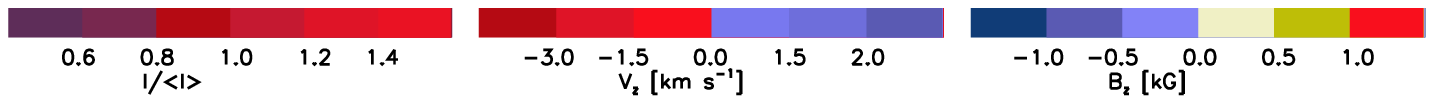}
              }
           \caption {The same as in Figure~\ref{m_54} but for
the second MHD simulation.  Dashed lines mark the selected
region for Stokes diagnostics with absolute flux density of
140~G at $\log\tau_R=0$ (140-MHD).
              }
\label{m_16}
\end{figure}
\noindent

The  spatial-temporal evolution of the simulated
magnetogranulation in the second simulation with the initial
$<B>=1.6$~G is shown in Figure~\ref{m_16}. The BPs are not
seen in the continuum at $\lambda 500$~nm
(Figure~\ref{m_16}a). Why? According to the simulation only
weak-field fluxtubes are formed. It is believed that the BPs
are caused by lateral radiation leakage scattering from
deeper layers of the magnetic element. The magnetic field of
the weak-field fluxtubes is not strong enough for their
evacuation  in their low photosphere levels. We assume that
such conditions can occur in the solar photosphere too. A
detailed study of observed BPs can be found in
\inlinecite{Sanchez04}, \inlinecite{Wijn05}, and
\inlinecite{Berger07}. These authors find that BPs are common
in the quiet Sun.   According to observations of G-band BPs
\cite{Sanchez04} their density is 0.3~Mm$^{-2}$ while
\inlinecite{Wijn05} found  lower density,
0.02--0.05~Mm$^{-2}$. In addition, there are many BPs with an
intensity smaller than that of the mean photosphere
\cite{Sanchez04}. The BPs deficiency in our second simulation
contradicts the observations in this respect. 60 snapshots
are taken from second simulation in the range from 90  to
120~min (marked on Figure~\ref{m_16}). $B_z$ is varied from
$-1070$ to 1300~G. The mean unsigned field strength $<B>$ for
this sample is 252~G at the optical depth $\log \tau_R=0$.
The unsigned magnetic  flux density $<|B_z|>$ is 140~G. This
sample of  60 snapshots  called 140-MHD is used later in
Stokes diagnostics.

\section{Spectral Synthesis} 
\label{synthesis}

The Fe I $\lambda$ 1564.8 nm line is the most
Zeeman-sensitive spectral line in the solar spectrum because
of its largest Stokes-V peak separation. This separation is
unhampered by thermal line broadening effects, so that the
field strength measurement becomes basically model
independent if the magnetic field were constant.  Therefore
one can directly determine the field strength from V-peak
separations without the use of atmospheric models. A
limitation of this method is that this line is weakened in kG
magnetic concentrations more than other typical lines
\cite{Sanchez00}. To verify a diagnostic potential of this
line we compare  $B$ derived  from V-peak separations of the
synthetic  profiles  and original $B$ taken from the results
of simulations. We find satisfactory agreement  between them
for $B>200$~G. As to observations, infrared polarimetry
allows to measure the $V$ signals which are above the noise
level of 0.0002--0.0003 in units of continuum intensity $I_c$
(\eg, \opencite{Khom03}). This precision  is sufficient to
recover strong fields in observations with high spatial
resolution ($0.^{\prime\prime}5$ and better). Note that Fe I
1564.8 nm line is formed under the LTE conditions and its
temperature sensitivity  is low   (\eg, \opencite{Solanki92}). 
It is often used in  polarimetric
measurements (\eg, \opencite{Lin95}; \opencite{Meunier98};
\opencite{Khom03}) as well as for the analysis of MHD
simulation (\eg, \opencite{Shem03}; \opencite{Khom05a};
\opencite{Shelyag07}).

For our calculation we use the same line parameters as that
in the recent paper by \inlinecite{Shelyag07}. We adopt the
excitation potential of 5.43~eV, the oscillator strength of
$\log (gf)= -0.675$, and the iron abundance of $A_{Fe}=$~7.43.
Collisional broadening by neutral atoms is calculated
following \inlinecite{Barklem00}. We do not take into account
any micro- and macro-turbulent velocities. The line profiles
are computed with a spectral resolution of 5~m\AA~in the
wavelength range $\pm1.5$~\AA~around the line center  for
four Stokes parameters at the solar disc center in every
vertical column along the line-of-sight of every snapshot.
The solution of the Unno-Rachkovskii equations for the
polarized light is obtained by a fifth-order
Runge-Kutta-Felberg method. We use our SPANSAT code
\cite{Gadun88}  modified for the calculation of Stokes
profiles.  As a result we obtain 6272 Stokes-V profiles from
the 300-MHD run and 12~120 from the 140 MHD run. These two
samples of synthetic V profiles are used for Stokes
diagnostics  of simulated regions with different unsigned
flux densities.

Note that the observed Stokes-V profiles are strongly
affected  by noise. Therefore, when deriving the properties
of the synthetic line profiles the noise should be considered
too. In our special case the minimum of the mean amplitude of
synthetic Stokes-V signals is $ a_V=0.0007$ (a maximum is
0.15), while the noise level of Stokes-V observations is
lower (0.0002--0.0003, \opencite{Dominguez06b}; 0.0005,
\opencite{Lites02}). It follows that we  need not to take
into account noise in the analysis of the synthetic Stokes-V
profiles.

Together with  the Stokes profiles we calculate the
contribution and response functions described by
\inlinecite{Grossmann-Doerth88}. We find that the effective
formation depths of four Stokes profiles differ slightly from
each other. Previously they were studied in detail by
\inlinecite{Shem03}. The mean level of the effective
absorption in the photosphere for the Stokes-V peaks of the
1564.8~nm line changes from $\log\tau_5 =-0.4$ to
$\log\tau_5=-1.3$, depending on their formation place.
$\tau_5$ is the optical depth at $\lambda$~500~nm.
\begin{figure}

\centerline{\includegraphics[width=0.655\textwidth,clip=]{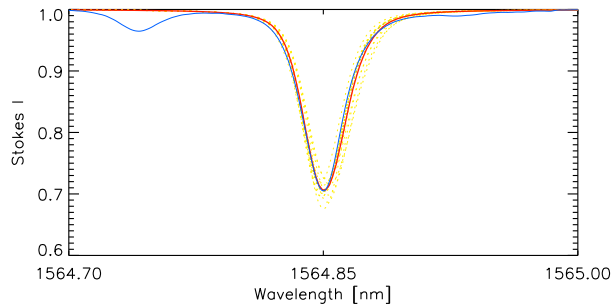}
              }
          \caption {Comparison of the spatially
averaged synthetic Fe~I~1564.8~nm line profile (shown in red)
and the observed quiet-Sun profile (blue) from the spectral
atlas of Delbouille, Roland, and Neven (1973). The red
profile is the average over the profiles (shown in yellow)
calculated from each of 12 snapshots (140-MHD run).
     }
 \label{prof_15}
\end{figure}
\noindent
In order to show how the simulated line  profiles agree with observations,
we show in Figure~\ref{prof_15} the
spatially averaged synthetic Stokes-I profile and the
observed spectrum of the quiet Sun \cite{Delbouille73}. The
mean synthetic Stokes-I profile is calculated  using 12 snapshots
with very weak magnetic fields. The
snapshots are taken from the 140-MHD run in 0.5 min in the
range from 90~min to 96.5~min. The number of snapshots is
optimized to exclude the effect of the 5~min oscillations.
The average synthetic and observed profiles agree
satisfactorily.

We also reduce the spatial resolution of the computed Stokes
profiles to the observed spatial resolution.  For this  we
perform a convolution of the synthetic Stokes profiles with the
point-spread function (PSF,  \opencite{Khom05a}). The PSF
represents a combination of an Airy function appropriate for the
German VTT and a Lorentz function describing the image degradation
caused by seeing. The PSF involves a 2-dimensional convolution
with the synthetic signals in the plane perpendicular to the LOS.
In the case of our 2D MHD simulation we have $V$ profiles which
change only along one spatial $X$-coordinate perpendicular to the
LOS. We assume that the profiles along the $Y$-coordinate
perpendicular to the LOS is identical to that of $X$-coordinate.
Thus each computed Stokes profile is individually convolved with
the PSF to be smeared in the plane perpendicular to the LOS. The
free coefficients of the PSF are selected so that the continuum
contrast at $\lambda$~1564.8~nm after smoothing matches the
observed contrast of 3\% \cite{Khom03}. Based on this contrast
agreement we assume that the resolution of the smoothed images of
magnetogranulation  is close to the observed resolution  of  about
$1^{\prime\prime}$ \cite{Khom03}.
\begin{figure}
 \centerline{\includegraphics[width=0.75\textwidth,clip=]{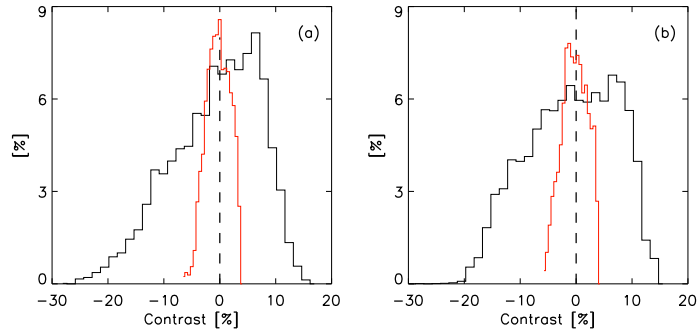}
              }
            \caption {Histograms of  continuum intensity contrast at
1564.8~nm. They were obtained for the 300-MHD (a) and 140-MHD
(b) regions before (black) and after (red) smearing synthetic
continuum intensity with the PSF.
             }
\label{contrast}
\end{figure}\noindent
\begin{figure}
 \centerline{\includegraphics[width=0.75\textwidth,clip=]{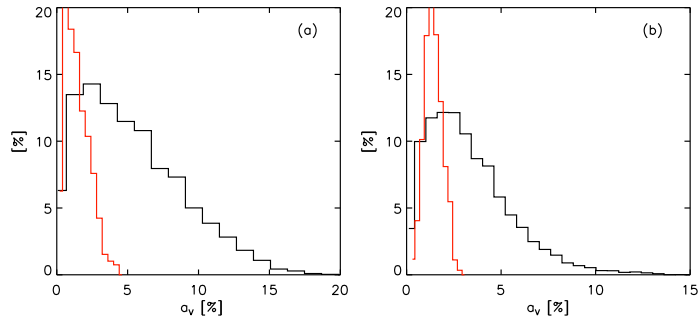}
              }
               \caption {Histograms of mean amplitudes ($a_V$) derived
               from the  synthetic Stokes-V
profiles  of the Fe~I 1564.8~nm line in the 300-MHD  (a) and
140-MHD (b) regions before (black) and after smoothing (red).
             }
\label{ampl}
\end{figure}
The synthetic rms contrast is about 8\%  before and about 3\%
after smoothing in the two simulated regions.
Figure~\ref{contrast} shows the contrast distribution
obtained before and after the spatial smoothing of the
continuum intensity. Note that the asymmetric shape of the
contrast distribution is more pronounced for the simulated
region with stronger average magnetic flux.

After smoothing the Stokes-V profiles change substantially.
The seeing transforms  the majority of the profiles to
broader and weaker ones. The amplitudes of red ($a_r$) and
blue ($a_b$)  wings and  their mean values,
$a_V=(a_r+a_b)/2$, decrease. The relative distribution of
$a_V$ is shown in Figure~\ref{ampl}. On the average, the
amplitudes of the synthetic Stokes-V signals $<a_V>~ =5.8$\%
(3.6\%) before and 1.7\% (1.5\%) after smoothing in the
300-MHD (140-MHD) region. They are one order of magnitude
larger than the observed $<a_V>$ in the internetwork regions
(\opencite{Khom03}; \opencite{Sanchez03}) and are closer to
that observed in magnetic network regions. This fact as well
as the larger absolute flux density in the simulated regions
(by a factor of 10) as compared to the quiet-Sun observations
implies that the simulated regions likely correspond   to a
network regions observed on the solar surface.

The number of Stokes-V profiles with irregular shapes
changes considerably after smoothing, specially
in the case of 300-MHD. We call the $V$ profiles
irregular if they have more than one zero-crossing.
The fraction of irregular synthetic profiles for 300-MHD
snapshots is 9\% before and  58\% after spatial smoothing.
For 140-MHD snapshots it is 12\% and 5\%, respectively. The
number of irregular Stokes-V profiles increases after
smoothing for 300-MHD, but it decreases for 140-MHD.  Why do
the two simulations behave so differently?  Since neighboring
$V$ profiles  influence each other  in the act of smoothing, we
considered the shape of  many synthetic profiles in detail
before and after smoothing. It turns out that magnitude and
sign of amplitudes ($a_b$ and $a_r$) and especially
zero-crossing shifts differ dramatically  from a profile to a
profile in 300-MHD run. The reason is that the range of
magnetic field strength and vertical velocities is
considerably larger in the 300-MHD region  than in the
140-MHD region. Note that according to
\inlinecite{Dominguez06b}  34\% of the analyzed observed IR
profiles  have  irregular  shape.

The irregular V profiles can introduce additional errors in
the evaluation of  Stokes-V peak separation, zero-crossing
wavelength,  area and amplitude asymmetry. Therefore in our
analysis only the V profiles with regular shape are used,
i.e., with  a well-defined  zero-crossing.

\section{Results of Stokes Diagnostics of Two Simulated Regions}
 \label{Magnetic}

\subsection{The Magnetic Field Strength Distribution} 

How are magnetic fields distributed over the surface of the
Sun outside active centers? This question is a hot topic
in current research  (for reviews, see \opencite{Steiner03b};
\opencite{Dominguez06a}). The weak and strong magnetic fields
are  strongly mixed in the solar photosphere (\eg,
\opencite{Sanchez00}; \opencite{Berger04};
\opencite{Rouppe05}). The field lines are very tangled (\eg,
\opencite{Weiss02}; \opencite{Stein06}). Many attempts have
been made to deduce distribution function of magnetic field
on   the solar surface. However, there is no agreement
between the distribution functions obtained in different
papers (\eg, \opencite{Lin95}; \opencite{Lin99};
\opencite{Sanchez00}; \opencite{Collados01};
\opencite{Lites02}; \opencite{Khom03}; \opencite{Socas02};
\citeauthor{Dominguez03}, \citeyear{Dominguez03,
Dominguez06a}; \opencite{Gonzalez06}). If the  magnetic field
distribution on the solar surface is continuous in the range
from 0 to 2000 G, the existing discrepancies   can be
explained by the following reasons: 1) visible and IR lines
"see" different structures because of their different
magnetic sensitivity \cite{Sanchez00, Socas03, Khom05b}; 2)
different techniques use simple or complex atmospheric models
\cite{Dominguez06a}; 3) spatial resolution of the observed
Stokes profiles and cancellation effects are different
\cite{Khom07, Orozco07}; 4)~the  noise level in the
measurements of very weak Stokes-V signals  is varied
\cite{Bellot03}; 5) the difference of the formation depth of
different lines used in observations \cite{Gonzalez06}.
Besides, the variety of physical processes in different
regions leads to different magnetic structures and different
magnetic field distributions \cite{Steiner03b}.

Realistic simulation of solar magnetoconvection
(\opencite{Atroshchenko96a}; \opencite{Steiner96};
\opencite{Grossmann-Doerth98}; \opencite{Stein00};
\opencite{Vogler05}) has  revealed that the magnetic fields
in the photosphere are structured from the greatest to the
smallest details and  have mixed polarities
 at small scales. There are  a few studies  in which field strength
distributions have been obtained  from magnetoconvection simulations
\cite{Cattaneo99,Steinetal02, Steiner03a, Vogler05, Stein06}. 
We also use the 300-MHD and 140-MHD samples of snapshots from the 2D MHD
simulation by \inlinecite{Gadun99a} to determine the magnetic
field distribution.

The distribution of field strength  on the solar surface is
usually characterized by a probability density function, or
shortly PDF,  (\eg, \citeauthor{Dominguez03},
\citeyear{Dominguez03, Dominguez06a}; \opencite{Steiner03b};
\opencite{Vogler05}). According to \inlinecite{Steiner03a}
the probability density function  $p(B)$  is defined by a
probability distribution function,
$$
 P(B)= \int_0^{B} p(B)dB,
$$ \noindent
which gives the probability of finding a
magnetic field strength less than $B$ at given location
within the given field of view.  Where $p(B)dB$ is the
probability of a magnetic field strength  to have at a given
location a value in the interval $[B,B+dB]$. The function
$p(B)$  is always positive and normalized to unity
$$
  \int_{-\infty}^{+\infty} p(B)=1.
$$
\noindent
Its first moment,
$$
 <B> = \int_0^\infty B p(B)dB,
$$
\noindent
gives the mean value of the magnetic field strength. The
second moment,
$$
 <B^2> = \int_0^\infty B^2 p(B)dB,
$$
\noindent
provides the unsigned magnetic energy density $
<B^2>/8 \pi$. In our case  $p(B)$ is not continuous because
we have a finite number of grid cells, $N$, with $B_i$
values ($i=1, ..., N$). We  can obtain a discrete PDF of
magnetic field strength in each simulated region.
 \begin{figure}
 \centerline{\includegraphics[width=0.55\textwidth,clip=]{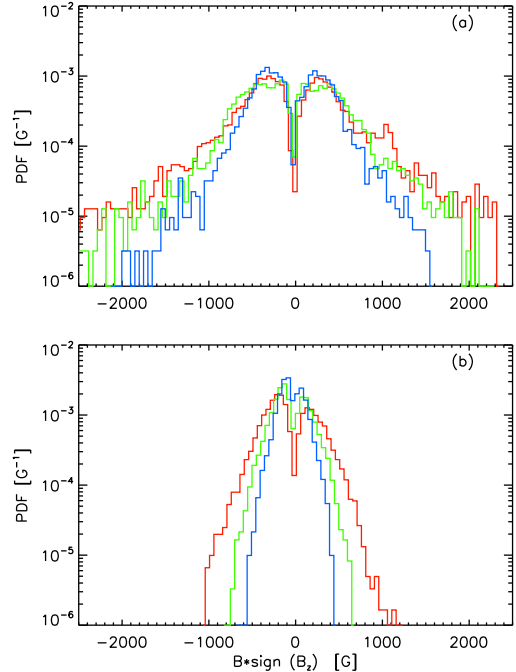}
              }
           \caption{Histograms of probability density function of
            the signed magnetic  field   strength in the 300-MHD  (a)
            and  the 140-MHD (b) runs at different photospheric levels
$\log \tau_5=0$ (red), -1 (green), and -2 (blue).
          }
\label{PDF_ful}
\end{figure}\noindent
For this purpose we divide whole range of the field strengths 
into equal bins of 50~G. Each individual
intrinsic field $B_i$ is assigned to one $k$th bin. The
discrete PDF for the $k$th bin, $[B_k, B_k+\Delta B]$, is
defined as $p_k=m_k/(N\Delta B)$, where $m_k$ is the absolute
frequency for the $k$th bin and  $\Delta B$ is the interval
of the bin.

Figure \ref{PDF_ful} shows histograms of the discrete PDFs of
signed magnetic  field  strengths in the two simulated regions.
They show the statistical properties of different
layers in the photosphere. The strongest magnetic fields
beyond 1 kG are located near $\log \tau_5 =0$.
They decrease with the height in the photosphere.

For comparison we can consider  PDFs derived from 3D MHD
simulation of magnetoconvection with different initial
conditions  for magnetic field (\opencite{Vogler05}, Fig.~5;
\opencite{Steinetal02}, Fig.~5).   \inlinecite{Vogler05} used
a vertical, homogeneous, unipolar, initial  field of 200~G.
\inlinecite{Steinetal02} simulated with no initial magnetic
field, but, lately horizontal magnetic fields were advected
into the simulated domain by upflows  across the lower
boundary. Our PDF (Figure~\ref{PDF_ful}a)  in shape more
closely resembles the PDF by \inlinecite{Steinetal02} than
the PDF by \inlinecite{Vogler05} because the initial magnetic
field of our simulation  and  of \inlinecite{Steinetal02} are
similar in appearance. As the simulation of
\inlinecite{Steinetal02} looks  the emergence of magnetic
flux through solar surface, the same is characteristic of our
simulation.

\begin{figure}
\centerline{
              \hspace*{0.015\textwidth}
               \includegraphics[width=0.4\textwidth,clip=]{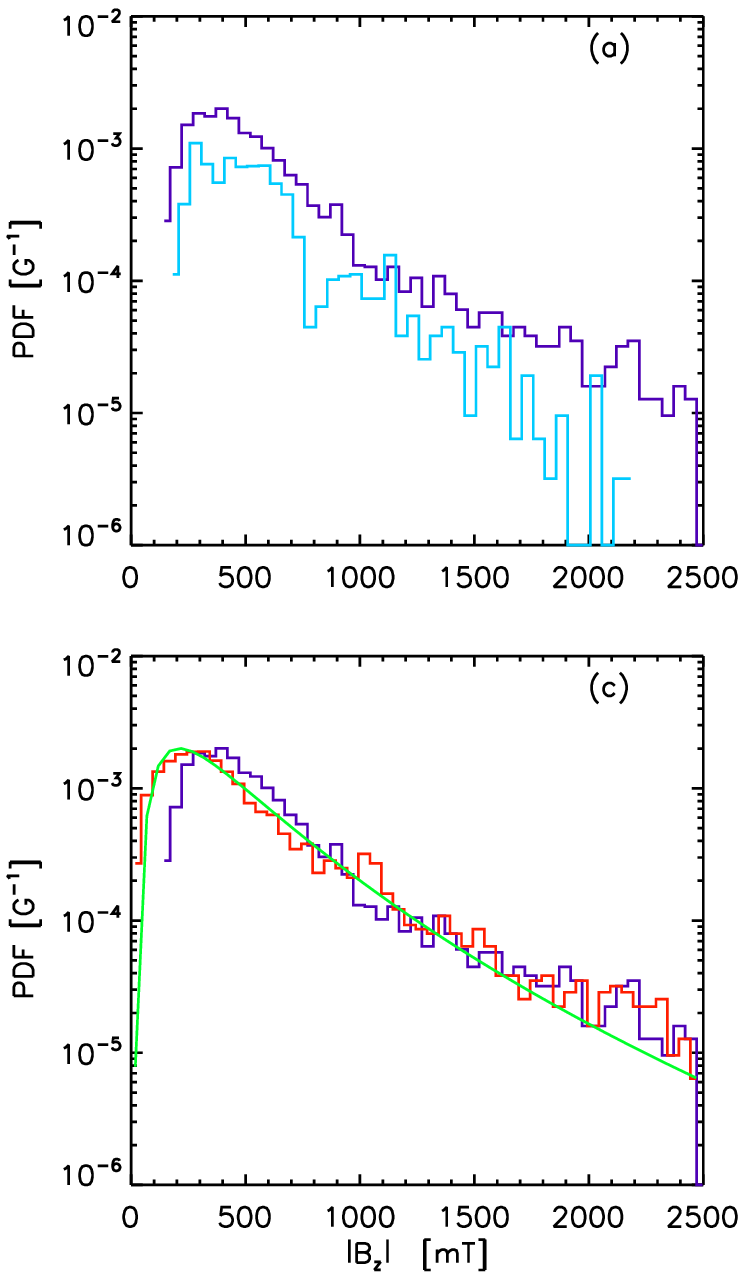}
               \hspace*{-0.03\textwidth}
               \includegraphics[width=0.4\textwidth,clip=]{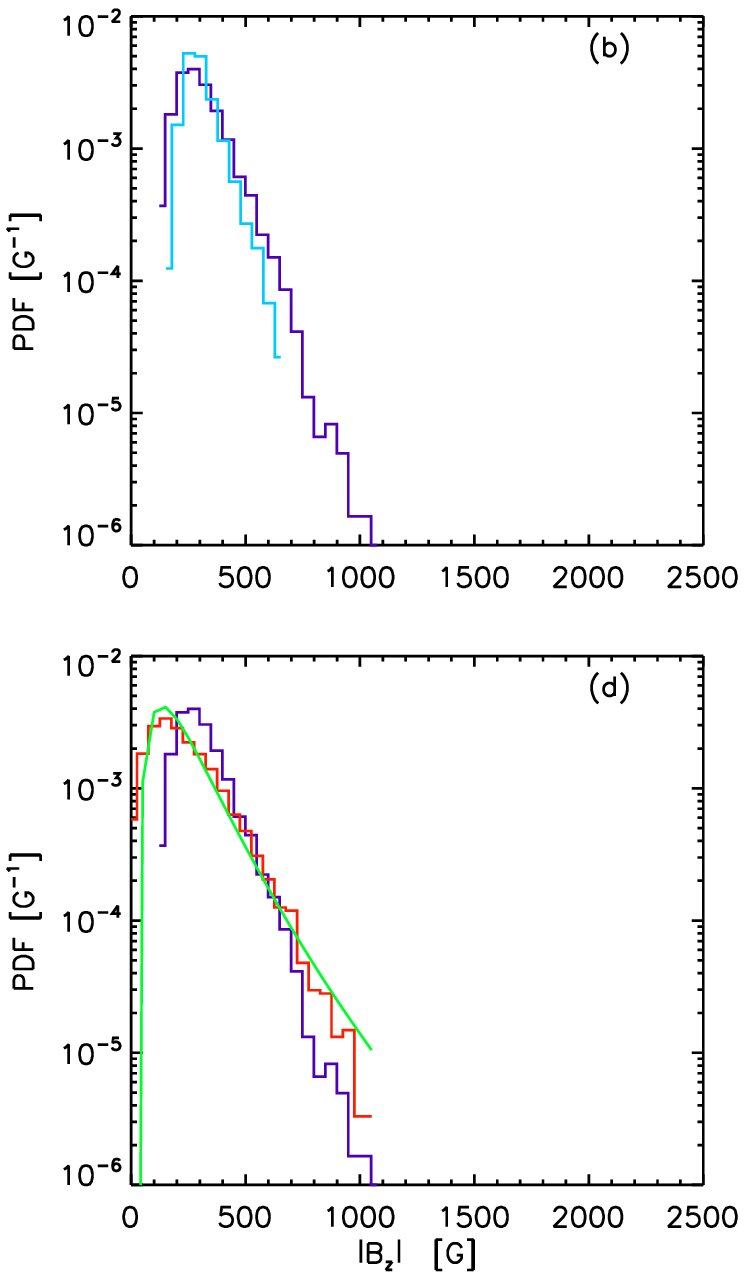}
              }

          \caption { 
          PDFs of the unsigned magnetic  field
          strength in the 300-MHD (a, c) and
          140-MHD (b, d)) simulated regions:
original (violet) and smoothed (blue) 15648-PDFs derived from
peak separations of the synthetic Stokes-V profiles  of the
Fe~I~1564.8~nm line without and with spatial  smoothing as
well as the MHD-PDFs (red) derived from the numerical data at
$\log \tau_5=0$, and its approximation by the lognormal
function (green). 
          }
\label{MHD-PDF}
\end{figure}

Just as the field strength PDF is derived from Stokes-V
observations, we obtain  the PDF  for simulated region from
the synthetic V profiles of the Fe~I $\lambda$~1564.8~nm
line. We use  the $V$ peak separations technique to determine
$B_i$ in each simulated element of each snapshot. The
inferred discrete PDFs of magnetic field strength in the
simulated regions are called 15648-PDFs. They are normalized
so that their integral is equal to 1 if all $N$ values of
$B_i$ of a given sample are considered. If only some portion
$n$ of all $B_i$ values is considered, the integral is equal
to $n/N$. Such normalization assumes that remaining portions
of the simulated region surface  is field-free. It does not
change the PDF's shape. Figure~\ref{MHD-PDF} (a, b) shows the
15648-PDFs based on the original (violet) and smoothed (blue)
Stokes-V profiles in the two simulated regions. The
difference between them increases for large field strengths.
The original (smoothed) 15648-PDFs derived from the two
simulated regions have $B_{max}$ about 330~G (270~G) and
about 230~G (250~G), respectively.   We can estimate the
reliability of  the $V$ peak separations technique comparing
the  PDFs derived from numerical simulation  and from
synthetic Stokes-V  profiles (original and smoothed). It can
be seen from   Figure~\ref{MHD-PDF} (c, d)  that the original
15648-PDFs and  the MHD-PDFs derived  from the simulations at
$\log\tau_5=0$ are in close  agreement in the range of field
strengths above 500~G. We conclude that the V-peak separation
method is reliable only for $B>500$~G.  The number of fields
obtained by this method  in the range 0--200~G is zero while
it is overestimated  in the interval 200--500~G. As a result
a maximum of the 15648-PDFs  is shifted with respect to a
maximum of the MHD-PDFs to stronger fields by 150~G.

Unfortunately,  we can not directly compare the 15648-PDFs or
the MHD-PDFs with the PDFs of magnetic field  strength
\noindent
observed in the network regions. When computing an observed
PDF  the knowledge of  individual filling factors for each
analyzed pixel is necessary. Since the filling factors are
different for weak and strong fields, taking them into
account can dramatically change both the shape of the
observed PDF and the magnitude of probability density. For an
example one can compare $|B_{app}|$ and true $|B|$ in
Figure~3 of \inlinecite{Lites02}.  Computation of the filling
factors and comparison between PDFs derived by different
techniques is not trivial task \cite{ Sanchez04a,
Dominguez06a}. On the other hand, the MHD-PDFs represent the
distribution of the field strength of each grid cell of a
specific  simulated region bounded by the sizes of
computation domain.

Let consider the shape of observed PDFs obtained by inversion
techniques which derive both true field strengths and their
filling factors. Figure~3 in the paper of
\inlinecite{Lites02} presents a histogram of relative
occurrence (which is proportional to the probability density
function) of absolute true field strength of magnetic network
and of their halo. This histogram reveals two peaks around
300--400~G (the halo fields) and 1300--1400~G (the strong
network elements). Using the other inversion technique,
\inlinecite{Socas02} found only a one maximum around kG
fields in the same network regions that were analyzed by
\inlinecite{Lites02}. The MHD-PDFs do not reveal clear
bimodal shape. Instead they have one very broad maximum
around 200--700~G for the 300-MHD run and 100--400~G for the
140-NHD run. Note the  MHD-PDF (300-MHD) in shape is
reminiscent of that found in observed quiet regions by
\inlinecite{Lin95} with a maximum around 300--700~G and
\inlinecite{Khom03} with a maximum about 350~G. The
MHD-PDF (140-MHD) is similar to the PDF derived by
\inlinecite{Orozco07} with a maximum close to 100~G.
Note, however, that in this comparison we are ignoring
the large fraction of field free photosphere inferred
from observations, which is not present in the
simulations. Observations show an unsigned flux
density one order of magnitude smaller than our simulations.

Figure~7 by \inlinecite{Dominguez06a}  compares the
semiempirical reference PDF of the quiet Sun with the PDFs
derived from a 3D MHD simulation by \inlinecite{Vogler05}.
The shape of all these PDFs is different from our MHD-PDFs.
The reference PDF for quiet-Sun region has $B_{max}=13$~G and
$<|B_z|>$~=~150~G.  The PDFs derived from three simulations
by \inlinecite{Vogler05} with initial field strengths of 10,
50, 200~G   have  $B_{max}$ about 5, 8, 50~G and
$<|B_z|>$~=~25, 100, 270~G, respectively. Instead, the
MHD-PDFs  for the 140-MHD, 300-MHD runs have $B_{max}=150$,
250~G and $<|B_z|>$~=~140, 300~G.  The most probable field
strength, $B_{max}$, in our two simulations is larger than in
the internetwork regions in  the reference PDF
\cite{Dominguez06a} and the simulations by \cite{Vogler05}.
The unsigned flux in the 140-MHD and 300-MHD is closer to the
unsigned flux in the 200~G simulation of  plages by
\inlinecite{Vogler05}. On the other hand, our two simulations
of magnetogranulation show magnetic features on the surface
level that are similar to elongated sheet-like structures
observed along intergranular lanes. They resemble the solar
magnetic network. Therefore it is not surprising that the
number of weak fields ($B<100$~G) in  our MHD-PDFs is less
than in the semiempirical reference PDF by
\inlinecite{Dominguez06a} obtained for the internetwork
fields of the quiet Sun.

Following \inlinecite{Dominguez06a}, we
calculate the fraction of simulated surface with
    $B > B^*$,
$$
 \alpha (B>B^*) = \int_{B^*}^\infty p(B)dB,
$$
\noindent
 the fraction  of unsigned magnetic field strength,
 $$
 \phi (B>B^*) =\frac{1}{<B>} \int_{B^*}^\infty B p(B)dB,
$$ \noindent
and the fraction of energy
 $$
 \epsilon (B>B^*) =\frac{1}{<B^2>} \int_{B^*}^\infty B^2p(B)dB.
$$ \noindent
We find that about 83\% of the surface area in
the simulated strong (weak) network is filled by the magnetic
fields of 100--1000~G (100--500~G). The fields with
$B>1000$~G occupy 9.7\% (0.3\%) of the surface area and
contribute with 55.2\% (0.6\%) of the magnetic energy.
The fields with $B>1500$~G have filling factors $\alpha
=4.2$\% (0) and contribute 32\% (0) of the magnetic energy.
The weakest fields ($B<100$~G) occupy a fraction
$\alpha=$~8\% (17\%) of the surface area  of the simulated
region and  contribute to the absolute magnetic energy with 0.2\%
(1.2\%).

In agreement with  \inlinecite{Dominguez06a}, we also find
that  the MHD-PDFs of unsigned field strength in the
simulated strong (weak) network at the level $\log \tau_5=0$
can be satisfactorily approximated by a lognormal
distribution $$
 p(B)= \frac{1}{\sqrt{\pi}\sigma B}\exp(-(\ln(B/B_0)^2/\sigma^2)
$$ \noindent with $\sigma = 1.0$ (0.8), and $B_0= 356$~G
(190~G). Figure~\ref{MHD-PDF}  (c, d, green lines) shows
$p(B)$ together with the MHD-PDFs. For comparison, the $
p(B)$ derived from the 3D MHD simulation by
\inlinecite{Vogler05} with an initial vertical field of 50~G
is characterized by  $\sigma = 1.7$, $B_0= 38$~G. These
values have been taken by us from \inlinecite{Dominguez06a}.
They differ from our values because of the different  flux
density of magnetic field in the computational domain as well
as different  initial conditions of the simulation.

\subsection{Magnetic Flux and Imbalance }

\begin{figure}
 \centerline{\includegraphics[width=0.65\textwidth,clip=]{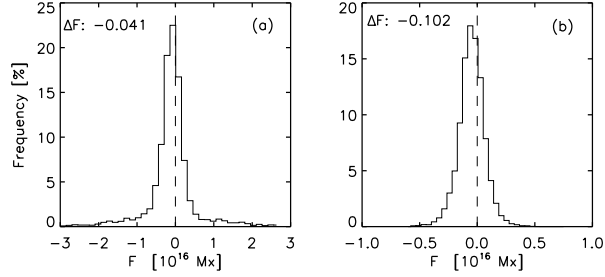}
              }
        \caption{Histograms of magnetic flux per resolution element
in   the 300-MHD (a) and 140-MHD (b) regions. $\Delta F $ is
the flux imbalance.
        }
\label{flux}
\end{figure}

We estimate the magnetic flux per resolution element as
$F_i=B_{z,i}S$, where $S$ is the area of a pixel, and
$B_{z,i}$ is the longitudinal magnetic field strength of
resolved $i$-element in  our 2D simulation at the level
$\log\tau_5=0$. Figure~\ref{flux} presents the histograms for
two simulated regions of the relative frequency   of $F_i$
within each bin. The maximum of these distributions is at
about $1\times 10^{15}$~Mx. The signed magnetic flux density
$<B_z>$ in the 300-MHD and 140-MHD regions is $-28$~G and
$-40$~G at $\log \tau_5=0$. The unsigned flux density
$<|B_z|>$ is $300$ and $140$~G, respectively.  It is close to
the values measured in actual observations of the magnetic
network \cite{Lites02} and  the ephemeral regions
\cite{Martin88}.
\noindent

Following \inlinecite{Lites02} we compute a flux imbalance
as:
$$
 \Delta F =  \frac{F^{+} + F^{-}}{|F^{+}| + |F^{-}|},
$$ \noindent
where $F^{+}$ and $F^{-}$ is positive and negative flux,
defined as the sum of $F_i$ fluxes directed towards and away
from the observer, respectively.

The imbalance is an important characteristic of the
magnetic flux distribution since it
shows the degree of mixture of positive and negative
fluxes. We find the imbalances in the
300-MHD and 140-MHD regions to be equal
 to $-0.041$  and $-0.102$. According to
\inlinecite{Lites02}, the flux imbalance for network regions
varies from 0.02 to 0.95 depending on the observed region.

\subsection{Line-of-sight Velocities }
 \label{Velocity}

\begin{figure}
 \centerline{\includegraphics[width=0.9\textwidth,clip=]{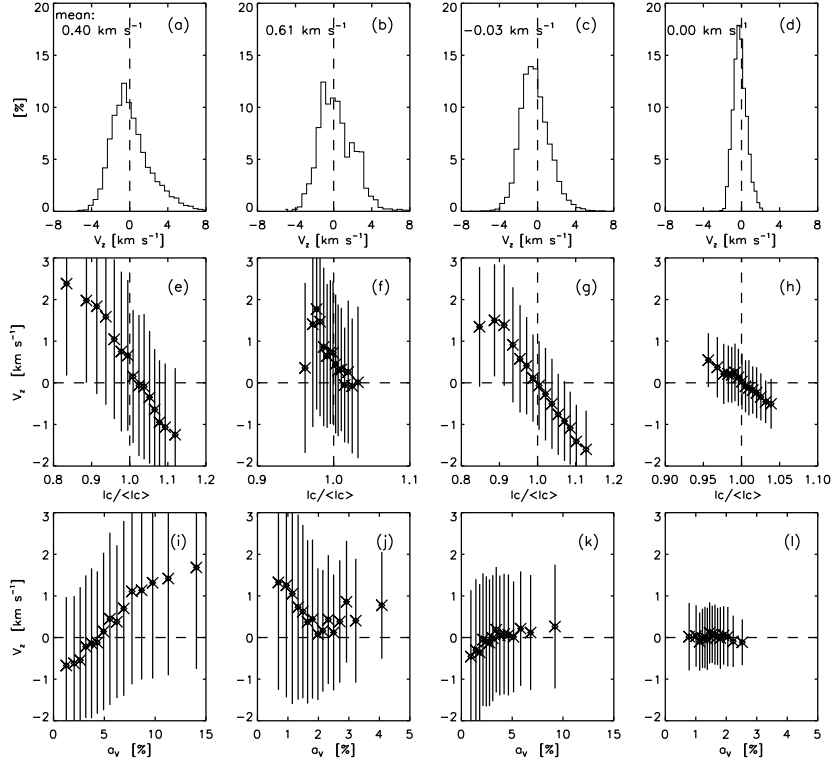}
              }
         \caption{Histograms of vertical velocity $V_z$,
and the statistical dependencies of $V_z$ on the
continuum intensity $I_c/<I_c>$ and the amplitude $a_V$
(indicator of magnetic flux density). The data are derived
from synthetic Stokes-V profiles before (a, e, i) and after
(b, f, j) their spatial smoothing in the 300-MHD region as
well as before (c, g, k) and after (d, h, l) their spatial
smoothing in the 140-MHD region. The $V_z$ values are binned
into a few intervals containing an equal number of points.
Error bars indicate the standard deviation of the individual
points within each interval.
         }
\label{Velos}
\end{figure}

By analogy with the measurements  of shifts of observed
Stokes-V, we measure the Doppler shifts of the synthesized
$V$ profiles of the Fe I $\lambda$ 1564.8~nm line relative to
the rest wavelength. Positive (red) shifts correspond to
downflows and negative (blue) shifts to upflows.
Figure~\ref{Velos} (a, c) displays the histograms of the
line-of-sight (LOS) velocities $V_z$ derived from the
zero-crossing shifts of the synthetic V profiles for two
simulated regions at the surface of $\log\tau_5=-1$.
The mean velocity is equal to about 0.40~${\rm km~s^{-1}}$
and $-0.03$~${\rm km~s^{-1}}$  for 300-MHD and for 140-MHD
regions, respectively. The width of the distribution ranges
from $-4$~${\rm km~s^{-1}}$  to 8~${\rm km~s^{-1}}$  and from
$-4$~${\rm km~s^{-1}}$  to 4~${\rm km~s^{-1}}$, respectively,
for these regions.  The larger the absolute magnetic flux
density in the simulated region, the stronger the downflows
in the magnetized plasma.

Figure~\ref{Velos} (e, g) shows the correlation between the
velocity and the continuum intensity contrast $I_c/<I_c>$ at
$\lambda$~1564.8~nm. $I_c$ is the continuum intensity in a
pixel of a snapshot and $<I_c>$ is the mean continuum
intensity in the snapshot. Similar to observations, the
regions of the simulated surface with the contrast more
than~1 are called granules and the rest of regions around the
granules are called intergranular lanes. As
Figure~\ref{Velos} suggests, the upflow velocity within the
granules of the 300-MHD (140-MHD) run is as large as about
$-1.2~(-1.8)~{\rm km~s^{-1}}$, on the average. We confirm the
observed granular blueshift of the Stokes profiles. In the
dark intergranular lanes, where fluxtubes are often located,
an average downflow velocity is as large as about 2.5
(1.5)~${\rm km~s^{-1}}$. This is the intergranular redshift
effect in the network regions. As to the observed
intergranular redshifts, it should be noted that no
significant systematic motions ($\pm 0.25$~${\rm km~s^{-1}}$)
inside fluxtubes are detected in spectropolarimetric
observations with spatial resolution worse
than~$1^{\prime\prime}$  (\eg, \opencite{Solanki86}). On the
other hand, a predominance of downflows is found in
observations with spatial resolution better
than~$1^{\prime\prime}$. In particular,
\inlinecite{Sigwarth99} obtain a mean redshift of about
0.5~${\rm km~s^{-1}}$ for magnetic structures  of the network
and the plages. \inlinecite{Langangen07}  find downflows of
about 0.2--0.7~${\rm km~s^{-1}}$ at the edges of magnetic
elements (bright points, ribbons, and flowers). In addition,
\inlinecite{Langangen07} also analyze realistic 3D
magnetoconvective simulations and find   downflows of
1.5--3.3~${\rm km~s^{-1}}$, i.e., it is much stronger than
observed.  The reason is that there are strong horizontal
gradients in the velocity field  on the small scales at the
periphery of fluxtubes.  The  observed redshifts are
underestimated due to insufficient spatial resolution. Our
simulated downflows of 1.5--2.5~${\rm km~s^{-1}}$  agree
well with the results of \inlinecite{Langangen07} and
confirm the intergranular redshift effect in the network
regions.

The dependence $V_z$ on $a_V$ in the 300-MHD region
(Figure~\ref{Velos}i  shows   that the large $V_z$ of the
downflows, on the average 2~${\rm km~s^{-1}}$, correspond to
strong magnetic fields. The upflows, with $V_z$ around
1~km~s$^{-1}$, correspond  to weak fields.
Figure~\ref{Velos}k  illustrates the downflows  in the
140-MHD region are  smaller than in the 300-MHD region. In
general, the  statistical dependencies confirm the tight
connection of the velocity field with small-scale magnetic
fields in the solar photosphere.

Besides, Figure~\ref{Velos} (b, f, j as well as d, h, l, p)
shows the velocity distribution and the statistical velocity
dependencies obtained after spatial smoothing of Stokes-V
profiles. The mean velocity is equal to 0.6~${\rm km~s^{-1}}$
(300-MHD) and 0~km~s$^{-1}$ (140-MHD). The larger the
absolute flux density, the stronger the changes due to
spatial smoothing. In particular, see the dependence of $V_z$~on~$a_V$.

Summarizing the results of our  LOS velocity analysis it is
necessary to point out that the mean LOS velocity  is greater
than zero within simulated strong network region with high
value of unsigned flux density. This is a result dominated
downflows which are appeared due to a formation of strong
magnetic concentrations with $B_z>1500$~G.  The correlation
between LOS velocities and the intensity contrast, and
Stokes-V amplitudes is in agreement with the physical
properties of convective motions in the photosphere. The
variations of these parameters reflect the changes of the
velocity field, temperature, and magnetic field  of the solar
plasma due to interaction of solar granules and small-scale
magnetic fields in the observed photosphere. Granular flows
and overshooting convection act to gather magnetic  field
lines  and to redistribute magnetic flux on the solar
surface. A fundamental property of the thermal convection in
a stratified photosphere  is the topological asymmetry
between hot upflows and cool downflows. Until the magnetic
field remains to be frozen in the plasma, i.e., until the
field reaches the condition of equipartition, the magnetic
flux is expelled from the granule interior to its boundaries,
the intergranular lanes, and also the upper photospheric
layers. The expulsion process may give rise to magnetic field
strength up to 1000 G at the radiating layer level in the
region of downflows $\cite{Gadun00}$. Hence, downflows,
strong magnetic field strength, and low temperature in the
photosphere  are related to each other. A similar statement
is true for upflows, weak magnetic field strength, and high
temperature.

\subsection{Asymmetry of Stokes-V Profiles of the Fe I 1564.8~nm Line }
\label{Asymmetry}
\begin{figure}
\centerline{
              \hspace*{0.015\textwidth}
               \includegraphics[width=0.45\textwidth,clip=]{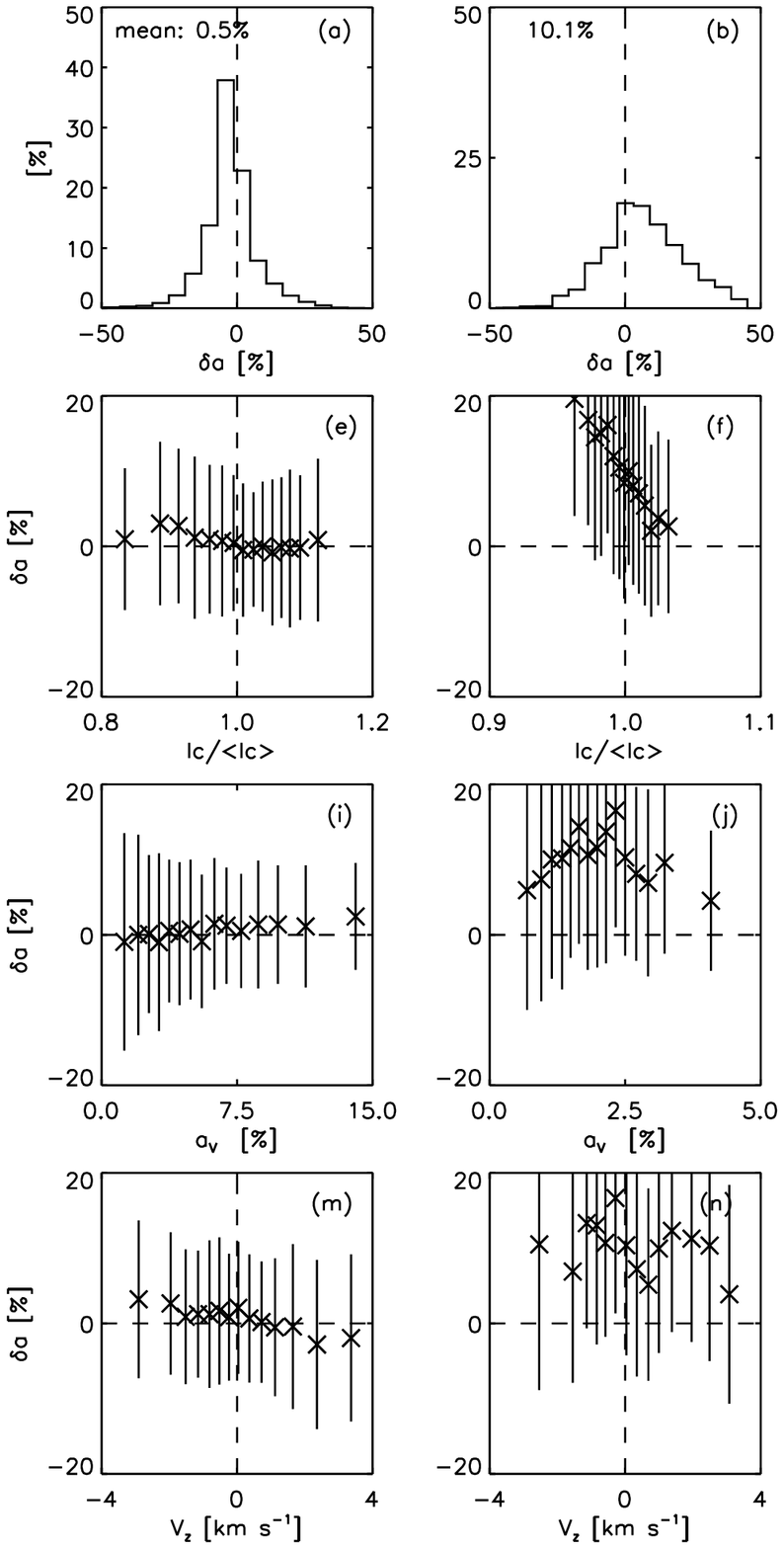}
               \hspace*{-0.01\textwidth}
               \includegraphics[width=0.45\textwidth,clip=]{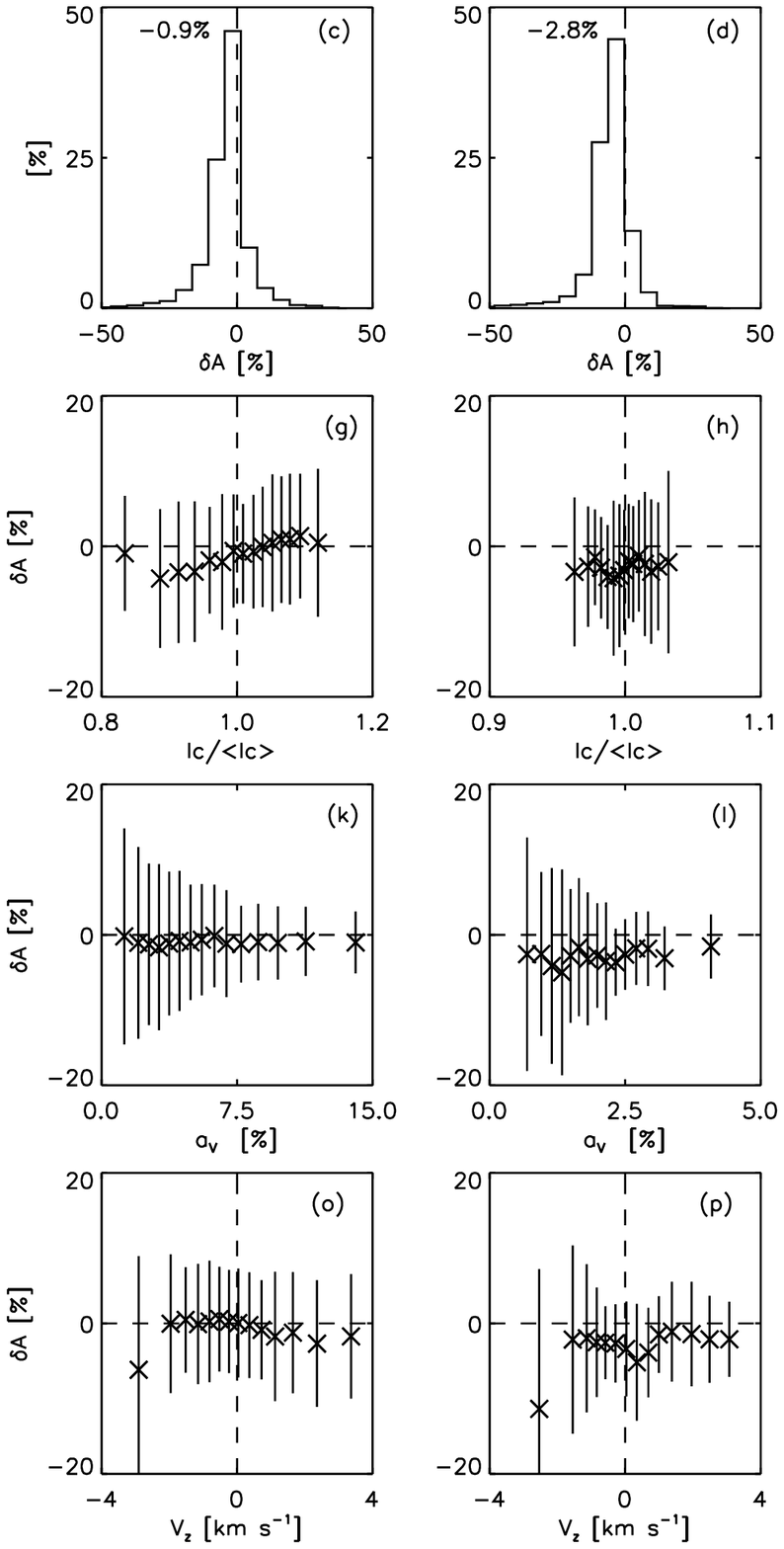}
              }
          \caption {Histograms of  amplitude  $\delta a$
and area   $\delta A$  asymmetries, and  their dependencies
on continuum intensity  $I_c/<I_c>$, amplitude $a_V$,  and
velocity $V_z$. These quantities are  derived from synthetic
Stokes-V profiles before (a, e, i, m as well as c, g, k, o)
and after (b, f, j, n as well as d, h, l, p) their spatial
smoothing in the 300-MHD region.
          }
\label{asym_54}
\end{figure}

The asymmetry of $V$ profiles is an indicator of the dynamic
processes in solar magnetized plasma.  According to
polarimetric observations (\eg, \opencite{Sanchez00};
\opencite{Sigwarth01}; \opencite{Socas02};
\opencite{Khom03}), $V$ profiles in active and quiet regions
on the Sun are variable in shape. Almost all of them are
asymmetric due to  gradients of the magnetic and velocity
fields along and across the line of sight \cite{Solanki93}.
Depending on the shape complexity, the profiles are said to
be regular (two lobes of opposite signs) and irregular (three
and more lobes or a single lobe).

We use the definition of the amplitude ($\delta a$)  and
area ($\delta A$) asymmetry of  \inlinecite{Solanki93}.
The amplitude and area asymmetry for the regular Stokes-V is
defined as:

$$
 \delta a =\frac{|a_b |-|a_r |} {|a_b |+|a_r |},~~~
 \delta A= \frac{|A_b |-|A_r |} {|A_b |+|A_r |},
$$

\noindent  
 where $a_b$, $a_r$, $A_b$, $A_r$ are the amplitudes
and areas of the blue and red wings.

\begin{figure}
\centerline{
              \hspace*{0.015\textwidth}
               \includegraphics[width=0.45\textwidth,clip=]{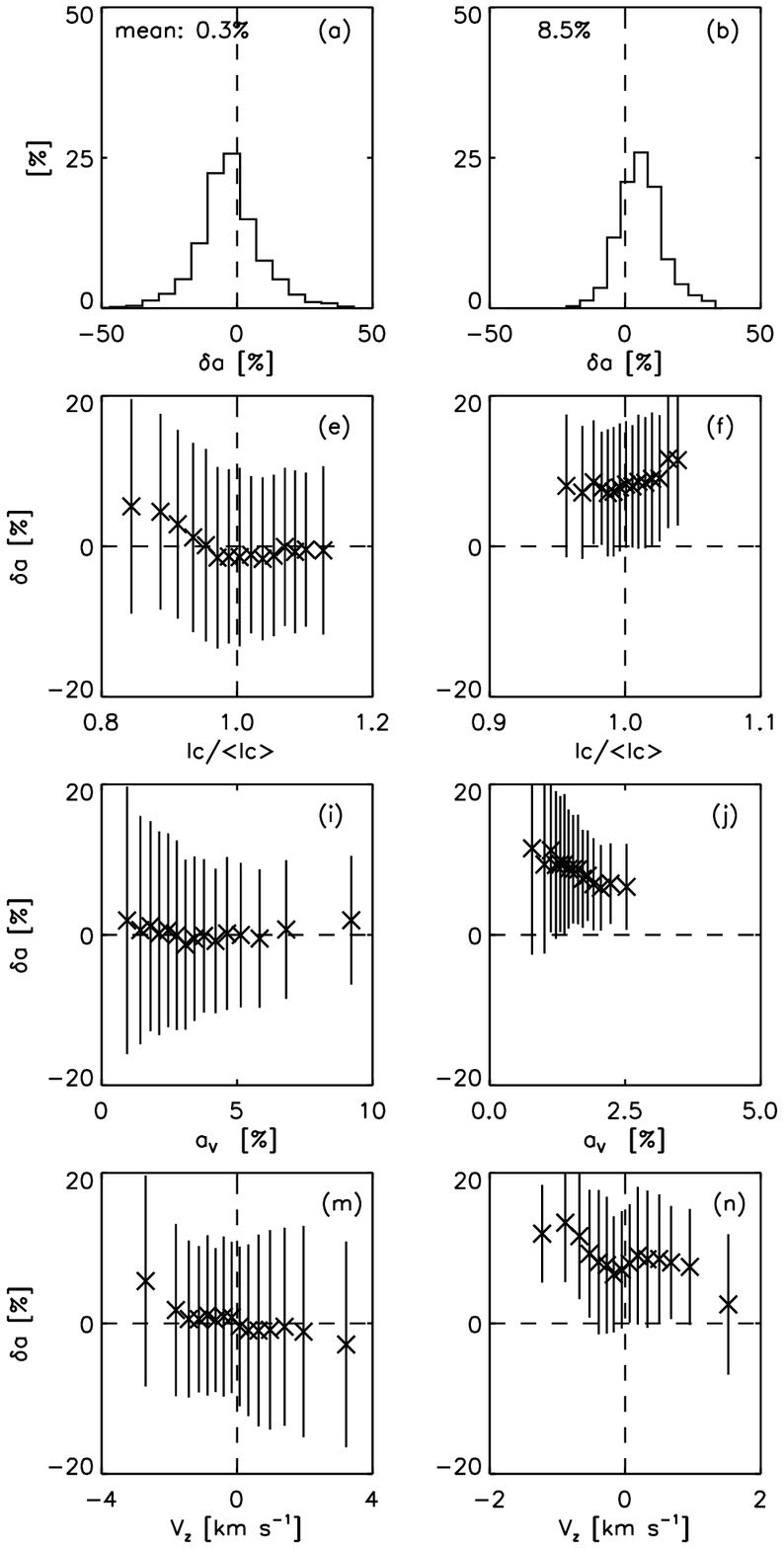}
               \hspace*{-0.01\textwidth}
               \includegraphics[width=0.45\textwidth,clip=]{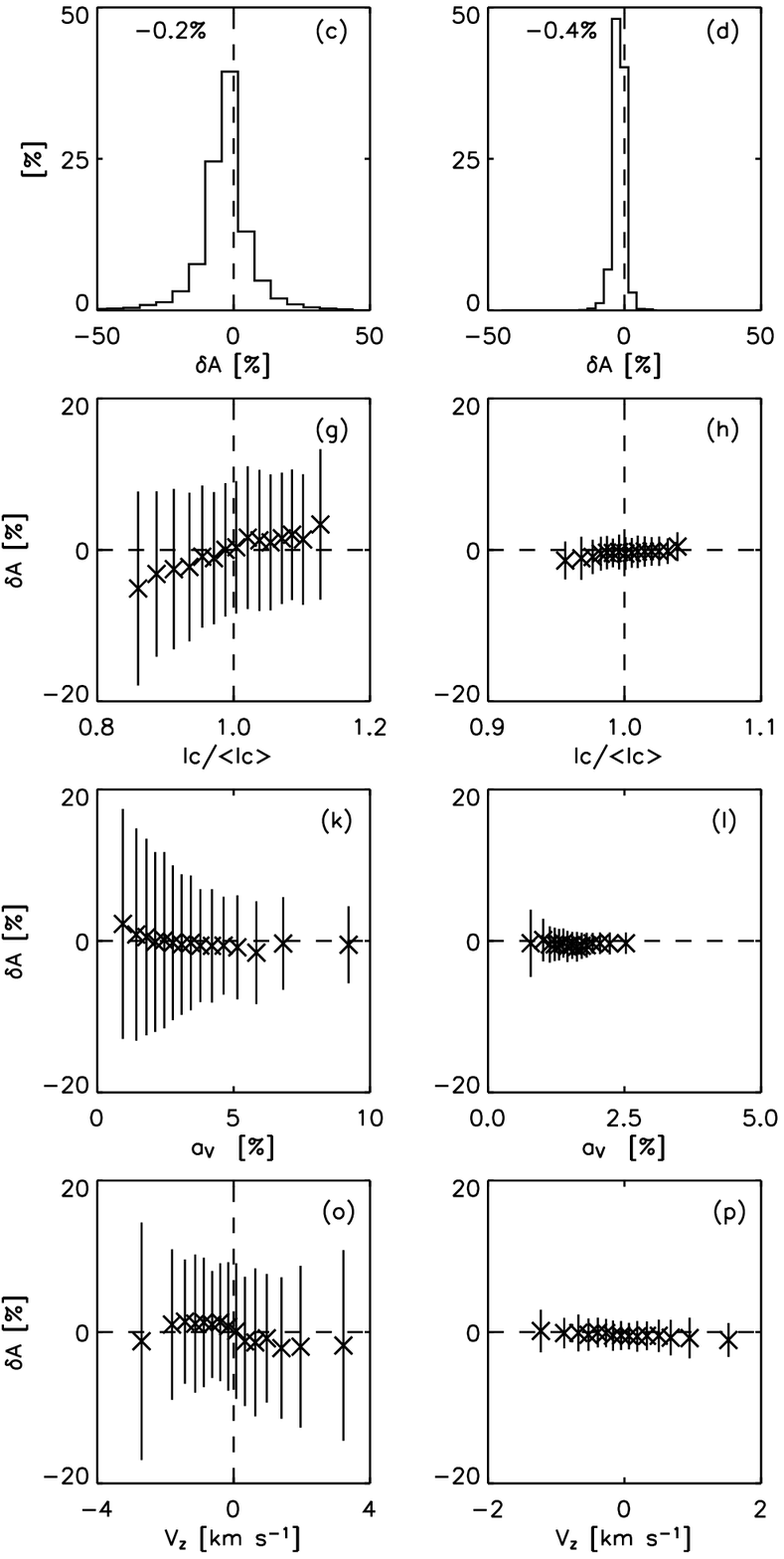}
              }
           \caption{The same as Figure~\ref{asym_54} but for the 140-MHD
region.
           }
\label{asym_16}
\end{figure}

The asymmetry of observed regular Stokes profiles has been
studied by many authors (\eg, \opencite{Grossmann-Doerth96},
\opencite{Martinez97}; \opencite{Sigwarth99};
\opencite{Khom03}). Previous and new results show that
outside sunspots  at the disc center the majority of regular
profiles have positive asymmetry of amplitudes and areas,
i.e., the blue wing is stronger than the red one; the
amplitude asymmetry is two times greater, on the average,
than the asymmetry of areas.  It was not clear for a long
time why the amplitude asymmetry exceeds the area asymmetry.

Figures \ref{asym_54} and \ref{asym_16} (a, c) show the
distribution of the amplitude and area asymmetries derived
for the 300-MHD and 140-MHD regions. The mean asymmetry of
synthetic V profiles in two simulated regions is very small:
$\delta a= 0.5$\% (0.3\%) and $\delta A=-0.9$\% ($-0.2$\%).

It is interesting to note that the character of  the
statistical dependencies of $\delta a$ and $\delta A$ on the
continuum intensity differ in the range $I_c/<I_c>~<1$
(Figures~\ref{asym_54}e and \ref{asym_16}g). If follows from
the V profiles formed within intergranules which have mainly
positive $\delta a$ and negative  $\delta A$. The magnitude
of $\delta a$ and $\delta A$  for these  V profiles is more
than for the profiles formed in the granules. The other
statistical dependencies ($\delta a$, $\delta A$  on $a_V$
and  $V_z$) do not show a clear tendency.

Spatial smoothing drastically affects amplitude asymmetry of
V-Stokes. The values of $\delta a$ basically become positive
and considerably larger for the majority of profiles. The
dependence of $\delta a$ on $I_c$/$<$$I_c$$>$ also varies significantly. 
Large $\delta a$ mostly occurs in the profiles formed in
intergranular lanes, where the effect of spatial smoothing is
the greatest. On the average, $\delta a=10.1$\% (8.5\%) and
$\delta A=-2.8$\% ($-4$\%) in the V-Stokes synthesized with
the 300-MHD (140-MHD) run. The spatial smoothing is one of
the main reasons of the  large observed amplitude asymmetry.
$\delta a$ is more
sensitive to the atmospheric effects than $\delta A$. This
may explain why the observed amplitude asymmetry  is by a
factor of two greater than the area asymmetry. Our results
demonstrate that the asymmetry of synthetic profiles (without
smoothing) is much smaller than that of smoothed ones. The
asymmetry of irregular synthetic Stokes-V profiles is
analyzed in detail by \inlinecite{Shem05}.

\section{Conclusion}  
\label{Concl}

We  use the realistic 2D MHD
simulations of nonstationary  magnetogranulation  by
\inlinecite{Gadun99a}, with initial bipolar magnetic
field, and with mean  unsigned field strength of 54~G and 1.6~G. Two
time sequences of snapshots with  an absolute flux density of
300~G and 140~G are analyzed in detail.
Most likely, the simulations reproduce magnetic
structures similar to the strong and the weak network observed on the
solar surface. Our results can be summarized as follows:\\

1. Based on the magnetic field strength probability density
functions,  PDFs, deduced from the two simulated
 regions, we find that   the PDF shape
depends on  the initial conditions of the magnetic field,
the absolute magnetic flux density in the given region, and
height in the photosphere.

2. According to the magnetic field  PDF obtained in the
simulated  strong network, the field strength varies from
1~G to 2800~G at $\log \tau_5=0$. The most probable field
strength is about 250~G. The mean unsigned strength is about
500~G. The unsigned flux density is 300~G and the signed flux
density is $-28$~G. The fields below 500~G fill most of the surface (70\%)
of given region. Very weak magnetic fields less than 100~G occupy 8\%
of the surface area, while the kG fields fill 9.7\%. The
filling factor of bright points with very strong kG fields
($B>1500$~G) is equal to 4.2\%.

3. The magnetic field strength in the simulated weak network
varies from 1~G to 1100~G at the level $\log \tau_5=0 $.
The most probable field strength is 150~G. The mean unsigned
magnetic field strength is about 250~G. The unsigned flux
density is 140~G and the signed flux density is $-40$~G. The
magnetic fields below 500~G occupy the major fraction (93\%)
of the  surface. Very weak magnetic fields less than 100~G
occupy 17\% and the kG fields fill 0.3\% of the  surface.

4. The magnetic flux is below $0.5 \times 10^{16} $~Mx in
90\% of resolved elements of simulated regions. Varying
magnetic polarities in the simulated regions produce a
polarity imbalance of $-0.10$ and $-0.04$.

5. The average line-of-sight velocity depends on the unsigned
flux density in the simulated region. The larger the magnetic
flux density, the stronger the downflows in the region. On
the average, the LOS velocities are  0.4~${\rm km~s^{-1}}$ in
the simulated strong network and 0~${\rm km~s^{-1}}$ in the
simulated weak network. The velocity scatter is in the range
from $-4$~${\rm km~s^{-1}}$ (upflow) to 8~${\rm km~s^{-1}}$
(downflow). On the average, the stronger downflows (about
2~${\rm km~s^{-1}}$) occur in integranular lanes filled by
strong magnetic field, while the stronger upflows (about
$-1.5~{\rm km~s^{-1}}$) occur in the granules. A close
correlation are found between LOS velocity and the granular
contrast as well as the magnetic field strength.

6. The amplitude and area asymmetries of synthetic Stokes-V
profiles do not exceed 1\%, on the average. If the V profiles
are spatially smoothed to $1^{\prime\prime}$,  the average
amplitude asymmetry rises to 10\% and  the average area
asymmetry is changed only slightly. Spatial averaging
influences the amplitude asymmetry by a larger extent
compared to  the area asymmetry. This may be a reason why
observed amplitude asymmetries are greater than area
asymmetries.

\acknowledgements This study was carried out thanks to the
great work of Aleksey Gadun who was the initiator of the
investigations of simulated solar  magnetogranulation and the
main executor of the MHD simulation. Unfortunately,  he will
never be able to continue the work started by  him  so
successfully. The author  thanks to A.~Gadun, S.~Solanki,
S.~Ploner for the productive collaboration, E.~Khomenko for a
code for the spatial smoothing of Stokes profiles. The author
is grateful to an anonymous referee for helpful comments and
important suggestions to improve the presentation of the
results and the content of the paper.

\end{article}

\end{document}